\documentclass[11pt,a4paper]{article}
\usepackage{amsmath,amssymb,amsthm,mathtools}
\usepackage[margin=1.25in]{geometry}
\usepackage[expansion=false]{microtype}
\usepackage{booktabs,array,graphicx}
\usepackage[colorlinks=true,linkcolor=black,citecolor=black,urlcolor=black]{hyperref}
\usepackage[round,authoryear]{natbib}

\newtheorem{proposition}{Proposition}
\newtheorem{lemma}{Lemma}
\newtheorem{corollary}{Corollary}
\theoremstyle{remark}
\newtheorem{remark}{Remark}

\newcommand{\dd}{\,\mathrm{d}}
\newcommand{\B}{\mathrm{B}}
\newcommand{\bvphi}{\bar\varphi}
\newcommand{\phic}{\varphi_{c}}
\newcommand{\Tstar}{T^{\star}}
\newcommand{\tstar}{t^{\star}}
\newcommand{\ustar}{u^{\star}}
\DeclareMathOperator{\Var}{Var}
\DeclareMathOperator{\Cov}{Cov}
\DeclareMathOperator{\E}{\mathbb{E}}
\newcommand{\Ibeta}{\mathcal{I}}                 
\newcommand{\Hyp}{{}_2F_1}                        

\title{From Alpha to Execution: Trading Horizons, Regimes, and Schedules under Concave Impact and Hurst Scaling} 
\author{Dnyanesh Kulkarni \thanks{\texttt{dpk52@cornell.edu}}\\{\small\textit{FMI Technologies LLC}}%
\and
Revant Nayar\thanks{\texttt{rnayar@fmitech.net}} \\{\small\textit{FMI Technologies LLC}}%
}
\date{October 2026}

\begin{document}
\maketitle

\begin{abstract}
Classical execution models assume linear impact, diffusive prices, and an
exogenous horizon. We study execution under concave power-law impact and
Hurst-scaled price fluctuations, and make two contributions.

First, we choose the execution horizon by maximizing the Sharpe ratio of the trade's realized P\&L. Because price moves during the execution window are part
of that P\&L, executing more slowly reduces its variance; and because impact is concave, executing more slowly can also lower its cost per share. Urgency comes
only from alpha that is realized while the order is being worked. The problem
is governed by the signs of two quantities: i. the duration elasticity of impact $a$, which under a
structural prior equals the \textit{impact-persistence gap} $\Delta_{IP} \coloneq \beta-H$, with $\beta$ the impact concavity parameter and $H$ the Hurst exponent, and can take either sign; and ii. the net urgency
$\nu$, the rate of alpha realization net of exposure relief. When $\nu$ keeps its sign over the window, an interior optimum
    exists, to leading order in $T/t_f$, only when both are positive;
    otherwise the order should be worked at the fastest or slowest
    admissible pace. In the interior regime, the first-order condition involves no impact coefficient, order size or volume, and the objective is flat: for a representative large-cap trade the window within $1\%$ of maximal Sharpe spans more than an order of magnitude.

Second, given the window, we derive the shape of the schedule by
    convex programming under an envelope cost; the schedule is optimal for this cost, and we quantify what that
robustness costs in Sharpe ratio. For any increasing per-share cost of delay $w(t)$, the KKT solution has the form
$\varphi\propto(\lambda-w(t))_+^{1/\beta}$. For a Hurst-scaled drift envelope,
$w\propto t^{H}$, the normalized schedule depends only on $(\beta,H)$, has a closed form in terms of the incomplete beta function, is front-loaded, and tapers smoothly to zero; it saves $10$--$17\%$ of the envelope cost relative
to a constant-rate schedule (such as TWAP) of equal duration. Whether it uses the whole window
depends on a single dimensionless robustness level, for which we give a
calibration. 
\end{abstract}
\clearpage

\begingroup
  \small
  \setcounter{tocdepth}{2}         
  \tableofcontents
\endgroup

\section*{Disclaimer} \textit{This document is provided for informational and educational purposes only and does not constitute investment advice, a recommendation, or an offer to buy or sell any securities, financial instruments, or investment strategies. 
The views expressed are subject to change without notice and may not reflect the opinions of affiliated institutions or entities.}

\clearpage

\section{Introduction}\label{sec:intro}
 
Every execution desk faces two decisions for a parent order of size $X$:
how long to take, and how to spread the order within that time. The
classical frameworks of \citet{AC2000} (AC) and
\citet{OW2013} (OW) remain the standard tools for the second
decision, but they leave the first one to the user. Both treat the
execution horizon $T$ as an input, and both assume that impact is
linear in trading rate and that prices diffuse. That makes the horizon
look like a trade-off between impact and price risk, with risk aversion
setting the balance. This is a poor description of how the horizon is
actually chosen. A portfolio manager trades to capture a forecast
return, so what matters is the Sharpe ratio of the trade's realized
P\&L, not cost and risk in isolation. And the impact and price dynamics
that set this Sharpe ratio depart from the classical assumptions in two
well-documented ways.
 
\paragraph{Concave impact:}
The average price displacement caused by a meta-order of size $X$ against market
volume $V$ follows, to good approximation, a concave power law
\begin{equation}
  I \;\propto\; \sigma\,(X/V)^{\beta},\qquad 0.4\lesssim\beta\lesssim 0.7,
  \label{eq:sqrtlaw}
\end{equation}
with $\beta=1/2$ the canonical square-root law
\citep{LilloFarmerMantegna2003,AlmgrenThum2005,Toth2011,Farmer2013}.
 
\paragraph{Fractional price scaling:}
Standard models assume diffusive prices ($H=1/2$ below). Empirically, return increments may
exhibit effective Hurst scaling
\begin{equation}
  \Var[S(t+\tau)-S(t)] \;\propto\; \tau^{2H}
  \label{eq:hurst}
\end{equation}
with $H\neq 1/2$ over execution-relevant horizons, arising from long memory in
order flow and incomplete liquidity replenishment \citep{LilloFarmer2004,Bouchaud2004,Bouchaud2009}.
Here $H$ is the scaling exponent of \emph{price increments}; it is distinct from
the roughness exponent $H\approx 0.1$ of \emph{log-volatility}
\citep{GatheralJaissonRosenbaum2018}.
 
\paragraph{Contributions:}
Our paper has two stages. Section~\ref{sec:horizon} decides \emph{how long to allow}
for the order, by maximizing the Sharpe ratio of the trade's realized P\&L. Two
features distinguish our treatment. We carry the duration elasticity of impact,
$a$, as a primitive whose sign is not assumed; and we account for the covariance
between execution-window price moves and the forecast return, which reverses the
naive intuition that longer execution adds risk.\footnote{The effect is particularly relevant for medium-frequency strategies, where \(T/t_f\) need not be negligibly small and a material fraction of the forecast may realize while the order is being worked.} The result is a regime
classification (Proposition~\ref{prop:regimes}) in which Hurst scaling acts mainly
as a regime selector; within a regime, its effect on the optimal window is smaller.
Section~\ref{sec:traj} decides \emph{how to spread the order within the window}. It does so under a separate cost criterion, so the schedule is optimal for that criterion rather than jointly with the window (Remark~\ref{rem:consistency}), as in AC. A
convex programme with concave impact and a per-share cost of delay yields a
closed-form participation schedule; with a Hurst-scaled drift envelope its shape
is universal in $(\beta,H)$ (Proposition~\ref{prop:envelope}). The schedule
tapers to zero and may finish before the window ends (Remark~\ref{rem:stopping}). Section~\ref{sec:properties}
collects comparative statics and testable predictions; Section~\ref{sec:conclusion}
concludes. For convenience of the readers, we have gathered our notation in Table~\ref{tab:notation}.
 
\begin{table}[h]\centering\small\setlength{\tabcolsep}{6pt}
\begin{tabular}{|l|l|l|}\toprule
Symbol & Meaning & Unit\\\midrule
$X$ & parent order size, in the direction of the signal & shares\\
$\rho$ & market volume rate & shares/time\\
$x(t)$, $v=-\dot x$, $\varphi=v/\rho$ & quantity remaining, trading rate, participation & shares, shares/time, --\\
$\sigma_1$ & volatility scale: $\Var[S(t+\tau)-S(t)]=\sigma_1^2\tau^{2H}$ & currency/share/time$^{H}$\\
$t_f$, $\mu_f$ & forecast horizon, expected return over $[0,t_f]$ & time, currency/share\\
$m(t)$, $\tau_\alpha$ & alpha realization profile, realization timescale & currency/share, time\\
$T$, $\bvphi=X/\rho T$ & execution window, average participation over it & time, --\\
$I(T)=AT^{-a}$ & impact at fixed $X$; $a$ = duration elasticity & currency/share\\
$\iota=I/\mu_f$ & impact as a fraction of alpha & --\\
$\nu$, $\varepsilon_V$ & net urgency, exposure-relief elasticity (Prop.~\ref{prop:foc}) & --\\
$C$, $\theta$, $\Theta$ & impact prefactor, envelope strength, $\theta/C(1+\beta)$ & --\\
$\tstar=\ustar T$, $\phic$ & end of trading, critical participation (Prop.~\ref{prop:envelope}) & time, --\\
\bottomrule\end{tabular}
\caption{Notation}\label{tab:notation}
\end{table}

\section{Classical Models and the Source of Trajectory Curvature}\label{sec:classical}
 
We consider a parent order of $X$ shares executed over $[0,T]$, with $x(t)$ the
quantity still to be executed and $v(t)=-\dot x(t)\ge0$ the trading rate.
``Adverse'' always means against the direction of the order. The classical papers
are written for the case of liquidation (selling); acquisition (buying) is the mirror image and only fixes sign
conventions. In Section~\ref{sec:horizon} the order is a buy in the direction of a positive signal.
 
\paragraph{Almgren--Chriss:}
AC minimize $J_{\rm AC}=\int_0^T[\eta\dot x^2+\lambda\sigma^2x^2]\dd t$ with
temporary-impact coefficient $\eta$ and risk aversion $\lambda$, subject to
$x(0)=X$, $x(T)=0$. The Euler--Lagrange equation $\ddot x=\kappa^2x$,
$\kappa^2=\lambda\sigma^2/\eta$, gives
$x(t)=X\sinh(\kappa(T-t))/\sinh(\kappa T)$. The trajectory is front-loaded and
reaches zero at the exogenous $T$ with a nonzero terminal rate
$X\kappa/\sinh(\kappa T)$. This derivative discontinuity means that trading ends abruptly.\footnote{One may recover a smooth exponential decay $e^{-\kappa t}$ in the $T\to\infty$ limit, but that is not a feature of the finite-horizon
solution.}
 
\paragraph{Obizhaeva--Wang:}
OW model a block-shaped book of depth $D$ that recovers at rate $\kappa_R$:
$\dot q=-\kappa_Rq+v/D$, with execution cost $\int q\,v\dd t$ plus the cost of any
block trades. Because the cost is linear in the control, the interior of the
optimal policy is a singular arc on which the displacement $q$ is held constant;
this requires the constant rate $v=\kappa_R D q$. The solution is a discrete block
at $t=0$, a constant interior rate (TWAP execution), and a discrete block at $t=T$ \citep{OW2013}.
The interior trajectory is linear, not hyperbolic as in AC.
 
\paragraph{Source of curvature:}
The hyperbolic AC trajectory is produced by the quadratic inventory penalty,
which turns the optimality condition into a linear second-order ODE. It is not a
consequence of the impact model: pure OW has no such penalty and thus produces a linear
interior. \citet{Almgren2003} generalizes AC to nonlinear impact but retains the
variance penalty and hence a second-order boundary-value problem. Section~\ref{sec:traj}
instead charges each fill a per-share cost of delay $w(t)$, which is linear in the
control. The optimality condition then becomes pointwise: the rate at time $t$ is
set by the marginal cost of delay at $t$ alone, and the sign constraint $v\ge0$
lets the rate taper to zero.

\section{The Sharpe-Optimal Execution Horizon}\label{sec:horizon}
 
\subsection{Setup}
 
The trader holds a signal with expected return $\mu_f$ over $[0,t_f]$ and buys
$X$ shares over a duration $T\le t_f$ (the subscript $f$ refers to \emph{forecast}). Prices follow\footnote{Equation \eqref{eq:price} should be interpreted as a local conditional description over execution-relevant horizons rather than as a global model of price dynamics. The function $m(t)=\mathbb E_0[S_t-S_0]$ represents the forecast-implied alpha-realization profile at trade inception, while the fractional term captures residual uncertainty through its scaling and covariance structure over the horizon of interest. Neither ingredient is essential in its specific benchmark form: our analysis allows a general alpha-realization profile $m(t)$, and the role of fractional Brownian motion is primarily to supply a tractable model of non-diffusive scaling. Thus, the framework is intended to apply more broadly to finite-horizon execution problems in which both expected return and residual risk evolve materially while the order is being worked.}
\begin{equation}
  S_t = S_0 + m(t) + \sigma_1 B^H_t,
  \label{eq:price}
\end{equation}
where $B^H$ is a fractional Brownian motion and $m$ is a non-decreasing alpha
realization profile with $m(0)=0$, $m(t_f)=\mu_f$. We consider two benchmark profiles that span useful limiting cases: \emph{event} alpha, realized at $t_f$ ($m\equiv0$ on $[0,t_f)$), and
\emph{linear} realization over $\tau_\alpha\le t_f$, $m(t)=\mu_f\min(t/\tau_\alpha,1)$.
The two are chosen for tractability rather than realism: linear is the simplest
profile satisfying $m(0)=0$, $m(t_f)=\mu_f$ and monotonicity, introducing a single
timescale $\tau_\alpha$ and nothing else. Real signal decay is more often concave
(front-loaded), and the two profiles bracket the urgency question rather than
describe actual realization dynamics. They are not limits of one
family: for $T<t_f$, linear realization with $\tau_\alpha=t_f$ forgoes
the fraction $g=T/(2t_f)$ of alpha, while event alpha forgoes none.
Linear realization covers the range from $\tau_\alpha\to 0$
(instantaneous alpha) to $\tau_\alpha=t_f$; event alpha is the separate
limit of profiles concentrated at $t_f$.

In this section, fills are spread uniformly over $[0,T]$. Section~\ref{sec:traj}
chooses the shape of the schedule under a different criterion, an
expected cost with a per-share cost of delay. The two stages together
are therefore a two-stage heuristic, not a joint maximization of the
Sharpe ratio over the fill density. The flatness of Section~\ref{sec:flat}
concerns the duration of a uniform schedule only. It does not bound the
effect of reshaping the fills, which changes both the impact paid and
the variance (Remark~\ref{rem:consistency}).

\paragraph{Trade P\&L:} The realized P\&L per share is
\begin{equation}
  \Pi(T) \coloneq S_{t_f} - \bar S_T - I(T),\qquad \bar S_T=\frac1T\int_0^T S_t\dd t,
\end{equation}

so the total trade P\&L is $X\Pi(T)$.
 
\paragraph{Impact and the duration elasticity $a$:}
At fixed order size $X$, we write
\begin{equation}
  I(T)=A\,T^{-a},\qquad a\coloneq -\left.\frac{\partial\ln I}{\partial\ln T}\right|_X .
  \label{eq:impactT}
\end{equation}
The elasticity $a$ is the central primitive of this section. Its sign, which is not
assumed \textit{a priori}, determines whether extending the execution horizon mechanically reduces or increases impact: indeed, $a>0$ implies that spreading the order reduces impact, $a<0$ that it increases it.
 
\begin{remark}[Structural prior for $a$]\label{rem:prior}
If impact scales with volatility over the execution window,
$I=C\sigma_T(X/\rho T)^\beta$ with $\sigma_T=\sigma_1T^H$, then
$A=C\sigma_1(X/\rho)^\beta$ and $a=\beta-H$. Under this structural prior, the
duration elasticity is exactly the \emph{impact-persistence gap},
$\Delta_{\rm IP}\coloneqq\beta-H$, and $a<0$ whenever prices are more persistent
than impact is concave, which is precisely the case that motivates a Hurst
framework. The prior is testable: meta-order studies report that impact at fixed
size depends only weakly on execution duration \citep{Toth2011,Zarinelli2015},
i.e.\ $a$ near zero, so every regime of Proposition~\ref{prop:regimes} below is
empirically relevant.

In general, we recommend estimating $\hat a$ directly from execution data, e.g.\
from the conditional dependence of log impact on log duration \emph{after}
controlling for order size, liquidity and execution conditions, rather than as
the difference of $\hat\beta$ and $\hat H$ estimated in separate studies.
\end{remark}
 
\subsection{Moments of the trade P\&L}
 
\begin{lemma}[Mean and variance]\label{lem:moments}
Let $g(T)=(\mu_fT)^{-1}\int_0^Tm(t)\dd t$ be the fraction of alpha forgone by
filling over $[0,T]$. Then
\begin{align}
  \E[\Pi] &= \mu_f\bigl[1-g(T)\bigr]-AT^{-a},\label{eq:mean}\\
  \Var[\Pi] &= \sigma_1^2V(T),\qquad
  V(T)=\frac{t_f^{2H+1}-(t_f-T)^{2H+1}}{(2H+1)\,T}-\frac{T^{2H}}{(2H+1)(2H+2)}.
  \label{eq:var}
\end{align}
$V$ decreases strictly from $V(0^+)=t_f^{2H}$ to $V(t_f)=t_f^{2H}/(2H+2)$, and for
$T\ll t_f$
\begin{equation}
  V(T)/t_f^{2H}= 1-H\,\tfrac{T}{t_f}-c_H\bigl(\tfrac{T}{t_f}\bigr)^{2H}+O\bigl((T/t_f)^2\bigr),
  \qquad c_H=\tfrac{1}{(2H+1)(2H+2)}.
  \label{eq:varexp}
\end{equation}
\end{lemma}
 
\begin{proof}
Equation~\eqref{eq:mean} is immediate. For~\eqref{eq:var}, expand
$\Var[B^H_{t_f}-\bar B^H_T]=t_f^{2H}+\Var[\bar B^H_T]-2\Cov(B^H_{t_f},\bar B^H_T)$ using
$\Cov(B^H_s,B^H_t)=\tfrac12(s^{2H}+t^{2H}-|t-s|^{2H})$. Direct integration gives
$\Var[\bar B^H_T]=T^{2H}/(2H+2)$ and
$2\Cov(B^H_{t_f},\bar B^H_T)=t_f^{2H}+T^{2H}/(2H+1)-[t_f^{2H+1}-(t_f-T)^{2H+1}]/((2H+1)T)$.
Monotonicity follows by differentiation (verified for $H\in(0,1)$).
\end{proof}

\noindent Note that at $H=\tfrac12$ the variance is exactly linear, $V(T)=t_f-\tfrac23T$, so
$\varepsilon_V=\dfrac{T/3}{t_f-2T/3}$. 
 
\begin{remark}[Why execution-window noise is not an independent source of risk]\label{rem:additive}
It is tempting to treat execution-window noise as a separate risk added in
quadrature to forecast risk, $\sigma_f^2[1+(T/t_f)^{2H}/(2H+2)]$, which increases
with $T$. However, this omits the covariance term: the fill price $\bar S_T$ lies on the
same path as $S_{t_f}$, and a later fill is exposed to less of the forecast-horizon
noise. With the covariance included, variance \emph{decreases} with $T$. It follows that under event alpha and $a\ge0$ the Sharpe ratio is increasing in
$T$ exactly, since then $\dd\ln\mathrm{SR}/\dd\ln T=a\iota/(1-\iota)+\varepsilon_V>0$:
the optimum is the slowest admissible execution. For $a>0$, an interior optimum requires
alpha realized during the execution window.
\end{remark}
 
\subsection{Exact optimality condition}
 
\begin{proposition}[First-order condition]\label{prop:foc}
Define $\iota=AT^{-a}/\mu_f$, the alpha-realization elasticity $\gamma=Tg'(T)$,
the exposure-relief elasticity $\varepsilon_V=-TV'(T)/(2V(T))\ge0$, and the net
urgency $\nu=\gamma-\varepsilon_V(1-g)$. On $\{\E[\Pi]>0\}$,
\begin{equation}
  \frac{\dd\ln\mathrm{SR}}{\dd\ln T}
  =\frac{\iota\,(a-\varepsilon_V)-\nu}{1-g-\iota},
  \qquad\text{so at an interior optimum}\qquad
  \boxed{\;\iota^\star=\frac{\nu(T^\star)}{a-\varepsilon_V(T^\star)}\;}.
  \label{eq:foc}
\end{equation}
\end{proposition}
 
\begin{proof}
$\dd\ln\mathrm{SR}/\dd\ln T=T\,\E[\Pi]'/\E[\Pi]+\varepsilon_V$ with
$T\,\E[\Pi]'=\mu_f(a\iota-\gamma)$ and $\E[\Pi]=\mu_f(1-g-\iota)$; rearrange.
\end{proof}

Write $\Delta(T)\coloneqq a-\varepsilon_V(T)$ for the \emph{effective duration
elasticity}: the elasticity of impact net of the exposure relief from filling
later. Then~\eqref{eq:foc} reads $\iota^\star=\nu(T^\star)/\Delta(T^\star)$, and
$\dd\ln\mathrm{SR}/\dd\ln T$ has the sign of
\begin{equation}
  F(T)\equiv\iota(T)\,\Delta(T)-\nu(T)
  \label{eq:F}
\end{equation}
wherever $\E[\Pi]>0$. Since $\iota>0$, a stationary point requires $\Delta$ and
$\nu$ to have the same sign. Since $\varepsilon_V\ge0$, $\Delta\le a$ always.

Condition \eqref{eq:foc} relates two observables, $\iota^\star$ and
$\nu(T^\star)/\Delta(T^\star)$, and needs no unobserved impact-model parameter:
$\nu$ and $\varepsilon_V$ depend only on $T/t_f$, $T/\tau_\alpha$ and $H$. The
optimal window itself still depends on $A$, and hence on $X$ and $\rho$, through
\eqref{eq:Tstar}. It can therefore be checked against transaction-cost data, where $\iota$
is observed, without calibrating an impact model. From~\eqref{eq:varexp},
\begin{equation}
  \varepsilon_V\simeq\tfrac H2\tfrac{T}{t_f}+Hc_H\bigl(\tfrac T{t_f}\bigr)^{2H},
  \qquad
  \nu\simeq\frac{T}{2\tau_\alpha}-\frac H2\frac{T}{t_f}-Hc_H\Bigl(\frac T{t_f}\Bigr)^{2H}
  \quad(\text{linear realization, }T\le\tau_\alpha).
  \label{eq:nu}
\end{equation}

For $H<\tfrac12$, the $T^{2H}$ term in \eqref{eq:nu} above dominates at short durations,
so $\nu<0$ below
\begin{equation}\label{eq:Tcross}
T_\times = t_f\left[\frac{2Hc_H}{t_f/\tau_\alpha - H}\right]^{1/(1-2H)}
\end{equation}
and $\nu>0$ above it.
 
\subsection{Regime classification}
 
\begin{proposition}[Regimes]\label{prop:regimes}
Let $[T_{\min},T_{\max}]$ be the admissible durations ($T_{\min}$ set
by a participation cap, $T_{\max}\le t_f$ by mandate), restricted to
$\mathbb{E}[\Pi]>0$. In general, $\mathrm{SR}$ increases where
$F(T)>0$ and decreases where $F(T)<0$, with $F$ from \eqref{eq:F}, so $T^\star$
is located by the sign changes of $F$ on this set. Suppose in addition
that $\nu$ has constant sign on the window and is nondecreasing where
positive. 
(For linear realization both hold to leading order, with $\nu$ from (11),
when $T_{\max}\le\tau_\alpha$ and $H\ge\tfrac12$. The exact $\nu$ can turn
down near $T\sim t_f$ when $\tau_\alpha$ is close to $t_f$: for
$\tau_\alpha=t_f$ and $H=\tfrac12$, $\nu=x(1-x)/[2(3-2x)]$ with $x=T/t_f$,
which peaks at $x=(3-\sqrt3)/2\approx0.63$. For $T>\tau_\alpha$,
$\gamma=\tau_\alpha/2T$ decreases, and for $H<\tfrac12$, $\nu$ changes sign
at $T_\times$.) Then, to leading order in $T/t_f$:

\begin{enumerate}
\item[(I)] $a>0$ and $\nu>0$: $\mathrm{SR}$ is single-peaked; $\Tstar$
  solves~\eqref{eq:foc}, or is the nearest endpoint if that solution lies
  outside the window.
\item[(II)] $a\ge0$ and $\nu<0$: $\mathrm{SR}$ is increasing; $\Tstar=T_{\max}$.
\item[(III)] $a\le0$ and $\nu>0$: $\mathrm{SR}$ is decreasing; $\Tstar=T_{\min}$.
\item[(IV)] $a<0$ and $\nu<0$: for $|a|<\min(1,2H)$, SR is single-troughed and
  $T^\star$ is the better endpoint; for $|a|>\min(1,2H)$, SR is single-peaked,
  with $T^\star$ interior or at the nearest endpoint. Under the structural prior
  the second case is empty, since $|a|=H-\beta<H<\min(1,2H)$.
\end{enumerate}
\emph{Boundary $a=0$:} Here $F=-\nu-\iota\,\varepsilon_V$ exactly, so
the leading term vanishes and the sign is set by $\nu$ and the
correction together. Case (III) applies only if $\nu>0$ on the whole
window. Where $\nu<0$ at short durations ($H<\tfrac12$, $T<T_\times$),
$\mathrm{SR}$ first increases and the optimum is interior, at
$\nu(T^\star)=-\iota\,\varepsilon_V$.

If $\E[\Pi]\le0$ throughout the window, the trade should not be executed at size $X$
(see capacity, \S\ref{sec:capacity}).
\end{proposition}

\begin{proof}
To leading order the numerator of~\eqref{eq:foc} is $\iota a-\nu$, 
with $\iota\propto T^{-a}$ and, by hypothesis, $\nu$ of constant sign and nondecreasing where positive. In (II) and (III)
its sign is fixed. In (I) it is strictly decreasing, so it changes sign at most
once, from positive to negative. In (IV) it equals $|\nu|-\iota|a|$, and, using the leading-order scaling $|\nu| \propto T^{\min(1,2H)}$,
$|\nu|/(\iota|a|)\propto T^{\min(1,2H)-|a|}$ is increasing, so it changes sign at
most once, from negative to positive. For $|a|>\min(1,2H)$ the ratio $|\nu|/(\iota|a|)\propto T^{\min(1,2H)-|a|}$ is
decreasing, so the numerator $|\nu|-\iota|a|$ changes sign at most once, from
positive to negative.
\end{proof}

\begin{remark}[Corrections]\label{rem:corrections}
Exactly, the numerator of~\eqref{eq:foc} is $(\iota a-\nu)-\iota\varepsilon_V$.
Exposure relief already enters $\nu$ at leading order; the remainder
$\iota\varepsilon_V$ is the product of two small quantities, with
$\varepsilon_V=O\bigl((T/t_f)^{\min(1,2H)}\bigr)$, and its size relative to the
leading term is $\varepsilon_V/|a|$. It has two effects. It shifts the interior
optimum: \eqref{eq:foc} holds with $a$ replaced by $a-\varepsilon_V(\Tstar)$. And it reclassifies a one-sided boundary layer
$0<a\lesssim\varepsilon_V$ from regime (I) to regime (III); for $a<0$ and $\nu>0$ it reinforces the leading-order sign. The expansion requires $T\ll t_f$ across
the window; near $T\sim t_f$, $\varepsilon_V=O(1)$ and Proposition~\ref{prop:foc}
should be used directly.

Since $a$ may be empirically close to zero (Remark~\ref{rem:prior}), classifying a
trade requires resolving $\hat a$ to better than the layer width $\varepsilon_V(T)$,
which is roughly $T/3t_f$ at $H=1/2$. Where this precision is unavailable,
Proposition~\ref{prop:foc} should be used directly.
\end{remark}

\begin{figure}[htbp]\centering
\includegraphics[width=0.72\linewidth]{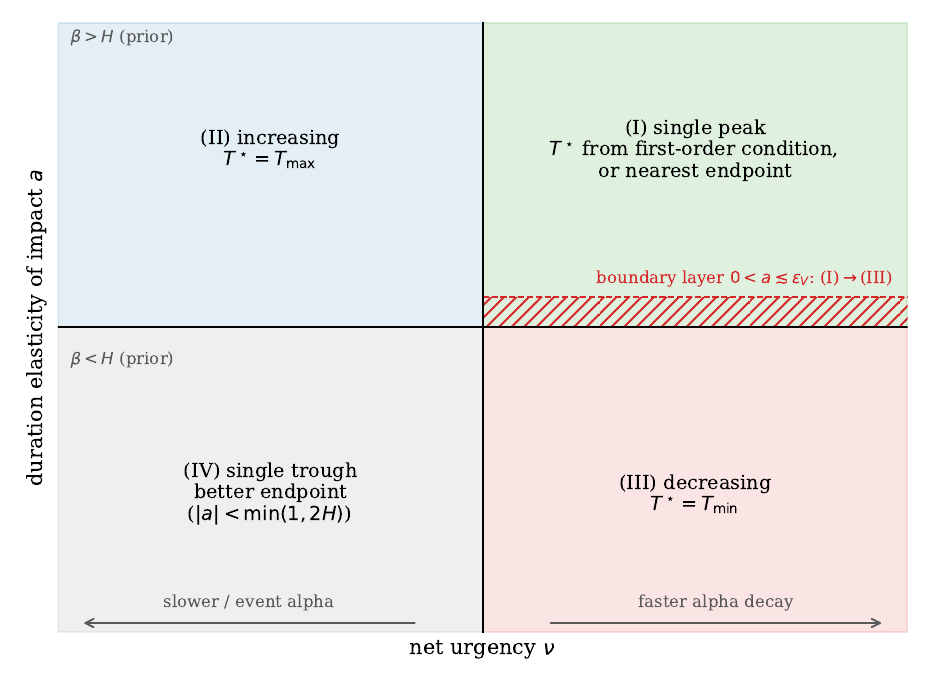}
\caption{Regimes of Proposition~\ref{prop:regimes} in the plane of net urgency
$\nu$ and duration elasticity of impact $a$, to leading order in $T/t_f$. Beyond
leading order, the hatched boundary layer $0<a\lesssim\varepsilon_V$ moves from
regime (I) to regime (III) (Remark~\ref{rem:corrections}); its width is
exaggerated in the figure. Under the structural prior, the upper half-plane is $\beta>H$.
Schematic: $\nu$ and $\varepsilon_V$ depend on $T$.}
\label{fig:regimeplane}
\end{figure}
 
The sign of $\nu$ is a race between alpha realization and exposure relief. With
linear realization, $\nu>0$ for small $T$ when $\tau_\alpha<t_f/H$ ($H>1/2$) or
$\tau_\alpha<\tfrac32t_f$ ($H=1/2$); both hold automatically since
$\tau_\alpha\le t_f$. For $H<1/2$, the $T^{2H}$ term dominates at short durations
and $\nu<0$ below $T_\times$ (eq. \eqref{eq:Tcross}): averaging
fills over anti-persistent noise reduces fill-price variance quickly, which
rewards slower execution at short durations.

\begin{figure}[htbp]\centering
\includegraphics[width=0.66\linewidth]{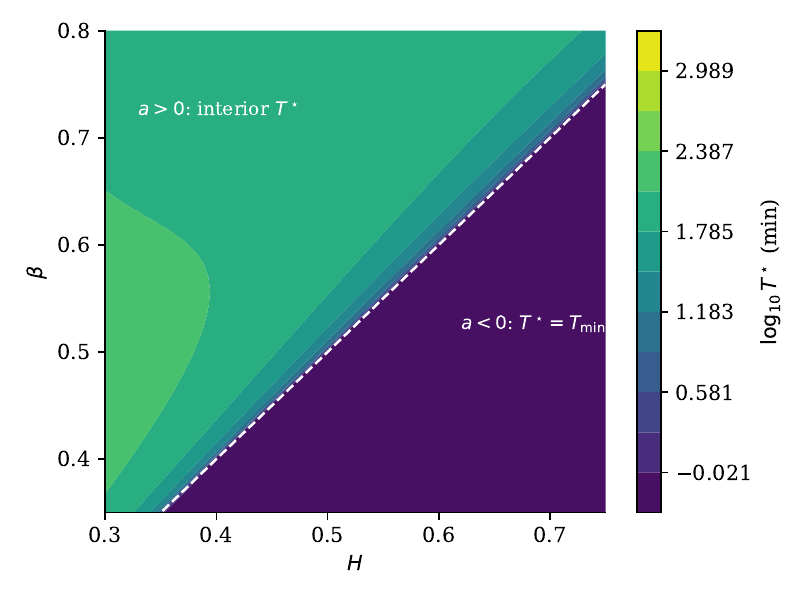}
\caption{Sharpe-optimal duration under the structural prior $a=\beta-H$, linear
alpha realization over $\tau_\alpha=t_f$, parameters of \S\ref{sec:amzn},
$T_{\min}$ at $20\%$ participation. Above the diagonal ($a>0$) the optimum is interior. Below it ($a\le0$)
the optimum collapses to $T_{\min}$, except in a thin strip along the
diagonal at $H\lesssim0.35$, where $\nu<0$ at short durations and the
optimum is interior (e.g.\ $T^\star\approx13$\,min at $\beta=H=0.3$;
Proposition~\ref{prop:regimes}, case $a=0$). The boundary-layer correction of Remark~\ref{rem:corrections} is below the resolution of the plot.}
\label{fig:regimes}
\end{figure}

Hurst scaling therefore enters in two ways: through $\nu$, which moves the optimum
within regime~(I) modestly, and, under the structural prior $a=\beta-H$, through
the sign of $a$, which selects the regime (Figure~\ref{fig:regimeplane}). The
second channel is the important one. The diagonal $\beta=H$ is the regime boundary
to leading order; beyond it, the boundary shifts by the layer width
$\varepsilon_V(T_{\min})$, about $10^{-4}$ for the example of \S\ref{sec:amzn}
and below the resolution of Figure~\ref{fig:regimes} (see also
Table~\ref{tab:amzn}).
 
\subsection{Closed form in the interior regime}
 
To leading order in $T/t_f$, \eqref{eq:foc} reduces to
$a\,AT^{-a}/\mu_f\simeq\nu(T)$. At $H=1/2$ with linear realization,
$\nu=c_\nu T$ with $c_\nu=1/(2\tau_\alpha)-1/(3t_f)$, and
\begin{equation}
  \Tstar\simeq\left(\frac{a\,A}{c_\nu\,\mu_f}\right)^{\!1/(1+a)}
  \;\overset{\tau_\alpha=t_f}{=}\;
  \left(\frac{6\,a\,A\,t_f}{\mu_f}\right)^{\!1/(1+a)} .
  \label{eq:Tstar}
\end{equation}
Volatility enters only through $A/\mu_f\propto\sigma_1/\mu_f$, alpha per unit of
short-horizon volatility. For general $H$ the implicit equation
$aAT^{-a}/\mu_f=\nu(T)$ with~\eqref{eq:nu} is solved numerically. The leading correction replaces $a$ by $a-\varepsilon_V(\Tstar)$ in the
coefficient (Remark~\ref{rem:corrections}).
 
\begin{remark}[Diversified books]\label{rem:diverse}
For one name in a diversified book, the relevant risk is the trade's marginal
contribution to portfolio variance, which scales down the exposure-relief term.
In the limit of negligible marginal risk, $\nu\to\gamma$ and
$\Tstar\to(2aA\tau_\alpha/\mu_f)^{1/(1+a)}$: diversification shortens the optimal duration, because it removes the
risk-based incentive to trade slowly and leaves only the impact-based one
($a>0$).\end{remark}
 
\subsection{Flatness}\label{sec:flat}
 
Differentiating (9) once more at the optimum, with
$p\coloneqq \mathrm{d}\ln\nu/\mathrm{d}\ln T$ ($p\simeq 1$ at $H=1/2$ with linear
realization, to leading order in $T/t_f$),
\begin{equation}
  \ln\mathrm{SR}(T)-\ln\mathrm{SR}(T^\star)\simeq
  -\tfrac12\,a(a+p)\,\iota^\star\bigl[\ln(T/T^\star)\bigr]^2 .
  \label{eq:flat}
\end{equation}
This is a local expansion, valid for $a>0$ and $a+p>0$. It drops the factor
$1/(1-g^\star-\iota^\star)$, a small-loss, small-exposure-relief approximation
($g^\star,\iota^\star,\varepsilon_V\ll 1$); for the AMZN example below, keeping it raises the curvature by about 12\%. The value $p=1$ is likewise leading order, not exact. Since $\iota^\star\ll1$ and $a$ is small, the curvature is tiny. We therefore report an indifference band rather than a point estimate (Figure~\ref{fig:flat}).
 
\subsection{Worked example: AMZN}\label{sec:amzn}
 
\begin{figure}[htbp]\centering
\includegraphics[width=0.74\linewidth]{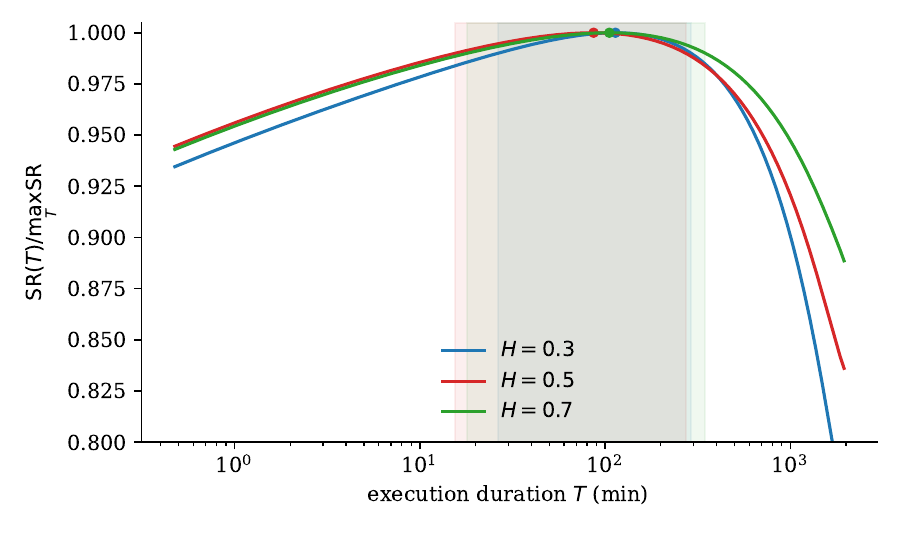}
\caption{$\mathrm{SR}(T)$ relative to its maximum, AMZN parameters, $a=0.1$.
Shading marks the $1\%$ indifference band for each $H$.}
\label{fig:flat}
\end{figure}

Consider\footnote{Market quantities such as the volume rate $\rho$ and the volatility scale $\sigma_1$ were estimated from AMZN trading on a single day in January 2026, so the example reflects realistic magnitudes for a large-cap stock. The remaining parameters are assumed for illustrative purposes.} $X=10^4$ shares, $\rho=1.05\times10^5$ shares/min, $\sigma_1=10.6$\,bp\,min$^{-1/2}$,
$t_f=1950$\,min, $\mu_f=19.2$\,bp, $\beta=0.6$, $C=1$, linear realization with
$\tau_\alpha=t_f$. Then $A=C\sigma_1(X/\rho)^\beta=2.59$\,bp\,min$^{a}$. At $H=1/2$
(thus $a=0.1$), \eqref{eq:Tstar} gives $99$\,min, its first-order
correction gives $86$\,min, and exact maximization gives $87$\,min, with
$\iota^\star=8.6\%$. By~\eqref{eq:flat} a factor-of-two error in $T$ costs
$0.2$--$0.3\%$ of Sharpe. The $1\%$ indifference band, computed by exact maximization rather than from
\eqref{eq:flat}, runs from 16 to 270\,min. It is asymmetric in $\ln T$: errors
toward shorter windows are cheaper than errors toward longer ones (Figure~\ref{fig:flat}). At $\Tstar=87$\,min, $\varepsilon_V\approx1.5\%$, so $a=0.1$ is comfortably
resolved from the boundary layer of Remark~\ref{rem:corrections}; a trade with
$a\lesssim0.015$ at this duration would not be. In the diversified limit of Remark~\ref{rem:diverse}, $c_\nu$ is replaced by $1/(2\tau_\alpha)$
and \eqref{eq:foc} gives 37 min in place of 99 min: a factor of 2.7, but still inside
the 1\% indifference band.
 
Table~\ref{tab:amzn} varies $H$. With $a$ held at $0.1$ (estimated directly),
$H$ moves $\Tstar$ by at most $30\%$, well inside the band. Under the structural
prior, $H\ge\beta$ switches the trade into regime~(III). Under event alpha $\nu=-\varepsilon_V<0$. Every row of the fixed-$a$
column, and the prior rows with $H\le\beta$, then sit in regime~(II)
with $T^\star=T_{\max}=t_f$. For $H=0.7$ the prior gives $a<0$, which is regime~(IV); $T_{\max}$ is the better endpoint for these inputs, but
this does not hold in general for $a<0$.
 
\begin{table}[h]\centering\small
\begin{tabular}{| c| c| c | c |}\toprule
$H$ & $\Tstar$, $a=0.1$ fixed & $\Tstar$, $a=\beta-H$ & regime (prior only)\\\midrule
0.3 & 114 & 134 & I\\
0.4 & 93 & 117 & I\\
0.5 & 87 & 87 & I\\
0.6 & 91 & $T_{\min}=0.5$ & III ($a=0$)\\
0.7 & 105 & $T_{\min}=0.5$ & III\\\bottomrule
\end{tabular}
\caption{Sharpe-optimal duration (minutes), AMZN parameters, exact maximization.}
\label{tab:amzn}
\end{table}

Note that in the fixed-$a$ column $T^\star$ is U-shaped in $H$. By \eqref{eq:nu}, the exposure-relief drag on $\nu$ is
$\tfrac H2\tfrac T{t_f}+Hc_H(T/t_f)^{2H}$: the linear term grows with $H$ while the
$T^{2H}$ term shrinks, so the drag is smallest near $H=\tfrac12$ and $T^\star$ is
smallest there.
 
\subsection{Capacity}\label{sec:capacity}
 
At fixed $T$, with $A=C\sigma_1(X/\rho)^\beta$ as in \S\ref{sec:amzn} (so
$A\propto X^\beta$ for any $a$), total expected P\&L $X\,\mathbb{E}[\Pi]$ is
maximized at
\begin{equation}
  X^{\max}=\rho\left[\frac{\mu_f(1-g)\,T^{a}}{(1+\beta)\,C\sigma_1}\right]^{1/\beta},
  \qquad I(X^{\max})=\frac{\mu_f(1-g)}{1+\beta}.
\end{equation}
At capacity impact consumes $1/(1+\beta)\approx 60\%$ of alpha. Capacity marks where
additional size stops adding P\&L; Sharpe-optimal trades sit far below it.
 
\section{Envelope-Optimal Execution Schedules}\label{sec:traj}
 
Section~\ref{sec:horizon} decided how long to allow for the order. Given that
window $[0,T]$ ($T=\Tstar$, or a mandated deadline), this section decides how to
spread the order within it. The answer is a schedule; whether the schedule happens
to use the whole window is a by-product (Remark~\ref{rem:stopping}).
 
\subsection{Cost functional}
 
With $\varphi=v/\rho$ we minimize
\begin{equation}
  J[v]=\int_0^{T}\Bigl[\kappa_I\,\rho\,\varphi(t)^{1+\beta}+w(t)\,\rho\,\varphi(t)\Bigr]\dd t,
  \qquad \int_0^{T}v\dd t=X,\quad v\ge0,
  \label{eq:J}
\end{equation}
with $\kappa_I=C\sigma_T=C\sigma_1T^H$. Here $T$ is the nominal window and $\kappa_I$ is held fixed at this
value throughout; in particular it is not re-evaluated at the time
$t^\star$ at which trading stops (Remark~\ref{rem:consistency}). The first term is local concave impact: a
fill at participation $\varphi$ pays $\kappa_I\varphi^\beta$ per share, and a
constant rate over the window reproduces the structural prior law $I=C\sigma_T\bvphi^\beta$ used in
Section~\ref{sec:horizon} (see Remark \ref{rem:prior}). The second term charges a fill at time $t$ a per-share
cost $w(t)$, increasing with $w(0)=0$: the price of having waited. Both terms are in
currency. There are two natural readings of $w$.
 
\emph{Alpha:} A share bought at $t$ rather than at $0$ forgoes expected return
$m(t)$. Taking $w=m$, $J/X$ is exactly the expected cost in~\eqref{eq:mean} for a
uniform schedule, so the mean part of Section~\ref{sec:horizon} is a special case of
this functional.
 
\emph{Risk envelope:} Rather than a variance penalty, which reintroduces a
dimensional risk-aversion coefficient, we charge a one-sided adverse move of
$\theta$ standard deviations accumulated by time $t$, consistent
with~\eqref{eq:hurst}:
\begin{equation}
  w(t)=\theta\,\sigma_1t^H .
  \label{eq:envelope}
\end{equation}
$\theta\in(0,1]$ is a dimensionless robustness level, the analogue of $\lambda$
in AC; it is not eliminated by the model (see Section~\ref{sec:theta}).
 
\begin{remark}[Fills versus inventory]
Integrating by parts, $\int w\,v\dd t=\int w'\,x\dd t$ when the order completes, so
a fill-based penalty is equivalent to an inventory holding cost at rate $w'$. An
inventory weight $(t/T)^Hx$ must carry a rate $1/T$ to have units of currency; it
corresponds to $w\propto t^{H+1}$ and changes the exponent, not the structure, of
what follows.
\end{remark}
 
\begin{remark}[The sign constraint is structural]
Because $w$ enters linearly, dropping $v\ge0$ would let the penalty credit
counter-trades and create spurious round trips. Thus, the constraint $v\ge0$ is part of the model and not merely a feasibility condition.
\end{remark}
 
\subsection{Existence, uniqueness, and the KKT schedule}

\begin{proposition}[Optimal schedule: existence, uniqueness, optimality]\label{prop:convex}
Let $w$ be continuous and non-decreasing on $[0,T]$ with $w(0)=0$. For $\lambda>0$ define
\begin{equation}
  \varphi_\lambda(t)=\left[\frac{\bigl(\lambda-w(t)\bigr)_+}{\kappa_I(1+\beta)}\right]^{1/\beta}.
  \label{eq:schedule}
\end{equation}
There is a unique $\lambda>0$ with $\rho\int_0^T\varphi_\lambda\dd t=X$, and
$\varphi_\lambda$ is the unique minimizer of $J$ over
$\mathcal F=\{\varphi\in L^{1+\beta}[0,T]:\ \varphi\ge0,\ \rho\int_0^T\varphi\dd t=X\}$.
In particular, every minimizer has this form.
\end{proposition}

\begin{proof}
\emph{Step 1: The multiplier.} Let $M(\lambda)=\int_0^T\varphi_\lambda\dd t$. Since
$w$ is bounded, $M$ is finite and continuous (dominated convergence). It vanishes
for $\lambda\le0$ and tends to $\infty$ as $\lambda\to\infty$. For $\lambda>0$ the
set $\{t:w(t)<\lambda\}$ contains a neighbourhood of $t=0$, and on it
$\varphi_\lambda$ is strictly increasing in $\lambda$, so $M$ is strictly
increasing on $(0,\infty)$. By the intermediate value theorem there is a unique
$\lambda>0$ with $M(\lambda)=X/\rho$. Fix it; then $\varphi_\lambda\in\mathcal F$
(it is bounded and non-negative).

\emph{Step 2: Pointwise minimization.} For each $t$, the function
$\ell_t(\varphi)=\kappa_I\varphi^{1+\beta}+(w(t)-\lambda)\varphi$ is strictly convex
on $[0,\infty)$ with derivative $\kappa_I(1+\beta)\varphi^\beta+w(t)-\lambda$. If
$\lambda>w(t)$ the derivative vanishes at $\varphi_\lambda(t)>0$; if
$\lambda\le w(t)$ it is positive for all $\varphi>0$. In both cases
$\varphi_\lambda(t)$ is the \emph{unique} minimizer of $\ell_t$ over $[0,\infty)$.
(These are the stationarity and complementary-slackness conditions: the derivative
equals zero where $\varphi_\lambda>0$ and is nonnegative where $\varphi_\lambda=0$.)

\emph{Step 3: Optimality.} For any $\varphi\in\mathcal F$, using
$\rho\int\varphi\dd t=X=\rho\int\varphi_\lambda\dd t$,
\begin{equation}
  \frac{J[\varphi]}{\rho}=\int_0^T\ell_t(\varphi(t))\dd t+\lambda\,\frac X\rho
  \;\ge\;\int_0^T\ell_t(\varphi_\lambda(t))\dd t+\lambda\,\frac X\rho
  =\frac{J[\varphi_\lambda]}{\rho}.
\end{equation}
All integrals are finite, since $w$ is bounded and $L^{1+\beta}[0,T]\subset L^1[0,T]$.
Equality forces $\ell_t(\varphi(t))=\ell_t(\varphi_\lambda(t))$ for almost every
$t$, hence $\varphi=\varphi_\lambda$ almost everywhere by the pointwise uniqueness in
Step~2. So $\varphi_\lambda$ is \textit{the unique} minimizer.
\end{proof}

Note that concavity of impact in \emph{size} does not conflict with convexity of cost in
\emph{rate}: the per-period cost $\varphi^{1+\beta}$ is convex because $1+\beta>1$.
 
\subsection{The envelope schedule}

Write $u=t/T$, $\Theta=\theta/[C(1+\beta)]$ and
\begin{equation}
  K(\beta,H)=\int_0^1(1-s^H)^{1/\beta}\dd s
  =\frac1H\,\B\!\left(\tfrac1H,\,1+\tfrac1\beta\right),
  \qquad \phic=\Theta^{1/\beta}K(\beta,H),
  \label{eq:phic}
\end{equation}
where $\B(p,q)=\int_0^1 s^{p-1}(1-s)^{q-1}\dd s=\Gamma(p)\Gamma(q)/\Gamma(p+q)$ is
the Euler beta function. $K$ is increasing in both $\beta$ and $H$.
 
\begin{proposition}[Envelope-optimal schedule]\label{prop:envelope}
Under~\eqref{eq:envelope}, with $\bar\varphi = X/\rho T$ and $\kappa_I$ fixed:
\begin{enumerate}
\item[(i)] If $\bvphi\le\phic$, the schedule is
  $\varphi(u)=\bigl[\Theta\,(\ustar{}^{H}-u^{H})_+\bigr]^{1/\beta}$ with
  $\ustar=(\bvphi/\phic)^{\beta/(\beta+H)}\le1$: trading tapers to zero at
  $\tstar=\ustar T$. If $\bvphi>\phic$, the schedule has the same form with $\ustar>1$, now fixed by
  $\int_0^1[\Theta(\ustar{}^H-u^H)]^{1/\beta}\dd u=\bvphi$; trading continues to $T$
  with a positive terminal rate.
\item[(ii)] If $\bvphi\le\phic$, then in the coordinate $\xi=t/\tstar\in[0,1]$
  \begin{equation}
    \frac{\varphi(\xi)}{\bvphi_{\rm act}}=\frac{(1-\xi^{H})^{1/\beta}}{K(\beta,H)},
    \qquad \bvphi_{\rm act}=\frac{X}{\rho\,\tstar}=\frac{\bvphi}{\ustar},
    \label{eq:shape}
  \end{equation}
  independent of $\theta$, $C$, $X$, $\rho$ and $\sigma_1$. The rate vanishes to
  order $1/\beta$ at $\xi=1$, so the landing is smooth.
\end{enumerate}
\end{proposition}
 
\begin{proof}
By~\eqref{eq:schedule}, $\varphi=0$ once $w(t)\ge\lambda$, i.e.\ for $t\ge\tstar$
with $\lambda=\theta\sigma_1\tstar{}^H$. Substituting into $\kappa_I=C\sigma_1T^H$
gives the stated $\varphi(u)$. The inventory constraint reads
$\bvphi=\int_0^{\ustar}\varphi\dd u=\Theta^{1/\beta}\ustar{}^{1+H/\beta}K$, which
gives $\ustar$; $\ustar=1$ defines $\phic$. Part~(ii) follows by substituting
$u=\ustar\xi$.
\end{proof}

For $\beta=1/2$ the inventory curve is an explicit combination of powers of $\xi$:
\begin{equation}
  \frac{x(\xi)}{X}=1-\frac{(H+1)(2H+1)}{2H^2}\,\xi+\frac{2H+1}{H^2}\,\xi^{H+1}
  -\frac{H+1}{2H^2}\,\xi^{2H+1}.
  \label{eq:traj}
\end{equation}
For general $\beta$, the curve is a regularized incomplete beta function, given in
Appendix~\ref{app:traj} together with its series and special cases.

\begin{figure}[htbp]\centering
\includegraphics[width=0.74\linewidth]{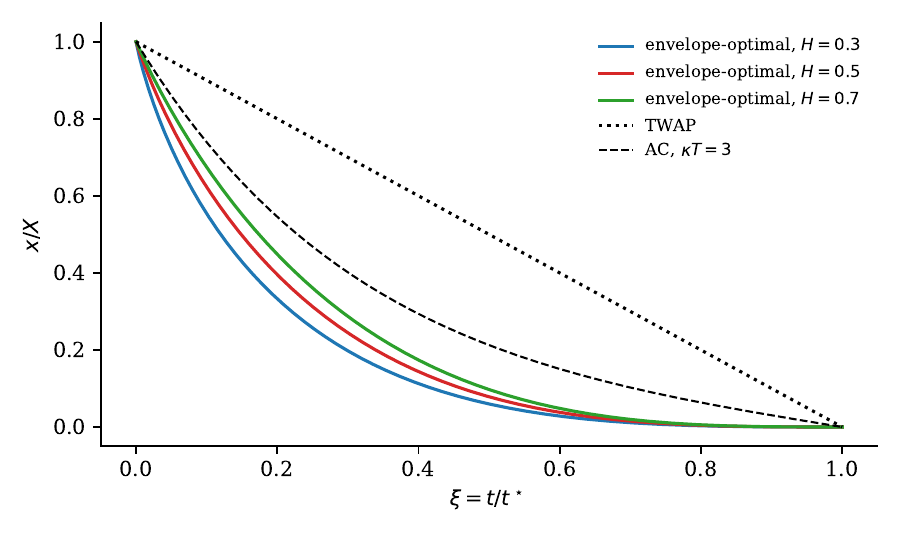}
\caption{Universal inventory curves~\eqref{eq:traj} ($\beta=1/2$) against TWAP and
an AC trajectory, each plotted over its own active period.}\label{fig:traj}
\end{figure}

The peak-to-average ratio over the active period is $1/K(\beta,H)$ ($=6$ at
$\beta=H=1/2$), and over the whole window is $1/(K\ustar)$
(Figures~\ref{fig:traj}--\ref{fig:part}). At $\beta=H=1/2$ and $C=\theta=1$,
$\phic=2/27\approx7.4\%$.
 
\begin{remark}[Where trading stops]\label{rem:stopping}
The closed form for $\ustar$ shows that the optimal schedule need not use the whole
window. We regard this as a corollary rather than a headline result for the following reasons. $\ustar$ is governed by $\theta$, which the model does not identify. For
AMZN with $\bvphi=0.11\%$, $\beta=0.6$, $H=1/2$: $\theta=1$ gives $\ustar=0.09$
($\tstar\approx8$\,min), $\theta=0.3$ gives $0.27$, $\theta=0.1$ gives $0.72$, and
any $\theta\le\theta_c\approx0.07$ uses the whole window. The property is also
generic: any increasing $w$ combined with $v\ge0$ yields $(\lambda-w)_+$, and
nothing in it is specific to Hurst scaling or concavity. Finally, where trading stops is not innocuous, for under the P\&L of Section~\ref{sec:horizon}
it is the main channel through which a pessimistic $\theta$ costs Sharpe
(Remark~\ref{rem:consistency}). In terms of AC, the difference is only at the endpoint: AC reaches zero at the
exogenous $T$ with $\dot x(T)\neq0$, while here the rate of trading tapers continuously and the
deadline need not bind. Only the shape ~\eqref{eq:shape} is robust to the choice of $\theta$.
\end{remark}
 
\begin{remark}[Two stages, one P\&L]\label{rem:consistency}

For a general fill density $p=v/X\ge0$ on $[0,T]$ with $\int p=1$, the
trade P\&L of Section~\ref{sec:horizon} has
\begin{align}
\mathbb{E}[\Pi_p] &= \mu_f-\int_0^T p(t)\,m(t)\,\mathrm{d}t
  -\frac{\kappa_I\rho}{X}\int_0^T\varphi(t)^{1+\beta}\,\mathrm{d}t,
  \nonumber\\
\operatorname{Var}[\Pi_p] &= \sigma_1^2\Big[t_f^{2H}
  +\iint p(s)p(t)R(s,t)\,\mathrm{d}s\,\mathrm{d}t
  -2\int p(t)R(t_f,t)\,\mathrm{d}t\Big],
  \label{eq:pnl-density}
\end{align}
with $R(s,t)=\tfrac12\big(s^{2H}+t^{2H}-|t-s|^{2H}\big)$ the fBm
covariance; uniform $p=1/T$ recovers Lemma~\ref{lem:moments}. Evaluating
\eqref{eq:pnl-density} on the schedules of Proposition~\ref{prop:envelope} shows that
schedule shape is not immaterial. For AMZN at $T=87$\,min
($\kappa_I=C\sigma_1T^H$), the envelope schedule attains $50\%$ of the
uniform optimum's Sharpe at $\theta=1$ ($t^\star=7.7$\,min, average
impact $10.5$\,bp against $1.7$\,bp), $79\%$ at $\theta=0.3$, $93\%$ at
$\theta=0.1$ and $96\%$ at $\theta=\theta_c$. The alpha reading $w=m$,
which matches the mean in \eqref{eq:pnl-density}, gives a nearly
uniform schedule and recovers the uniform Sharpe to within $10^{-4}$.
The envelope reading thus buys robustness against adverse paths at a
cost in expected Sharpe, and $\theta$ sets the exchange rate.
 
Throughout Section~\ref{sec:traj} the impact scale $\kappa_I=C\sigma_1T^H$ is
exogenous: it is fixed by the nominal window $T$ chosen in Section~\ref{sec:horizon}
and does not change if the schedule finishes at $t^\star<T$.
Proposition~\ref{prop:envelope} solves the problem only for fixed $\kappa_I$. Letting
$\kappa_I$ depend on the active horizon $t^\star$ defines a different
optimization, not a refinement of this one: for the AMZN inputs it
moves $t^\star$ from $7.7$ to about $1$\,min at $\theta=1$. Iterating
the two stages to a fixed point would not establish joint optimality
either. Maximizing $\mathbb{E}[\Pi_p]/\sqrt{\operatorname{Var}[\Pi_p]}$
over $p$ directly, a nonconvex ratio problem, is left open
(Section~\ref{sec:conclusion}).
\end{remark}

\subsection{Calibrating \texorpdfstring{$\theta$}{theta}}\label{sec:theta}
 
$\theta$ enters the schedule only through $\Theta$, and only $\ustar$ and the
initial rate $\varphi(0)=(\Theta\ustar{}^H)^{1/\beta}$ depend on it. Two calibrations
are available. (a) \emph{Shortfall matching:} choose $\theta$ so that the model's
predicted average shortfall matches TCA data. (b) \emph{Window use:} $\ustar<1$ if and
only if
\begin{equation}
  \theta>\theta_c=C(1+\beta)\bigl(\bvphi/K(\beta,H)\bigr)^{\beta}.
  \label{eq:thetac}
\end{equation}
For AMZN ($\Tstar=87$\,min), $\theta_c\approx0.07$, so a one-standard-deviation
envelope ($\theta=1$) is far more pessimistic than a $19$\,bp alpha justifies.
A parameter-free alternative for a risk-neutral trader is $w=m(t)$. With linear
realization over $\tau_\alpha$, the algebra above holds with $H$ replaced by $1$ and
$\theta\sigma_1T^H$ by $\mu_fT/\tau_\alpha$, where $K(\beta,1)=\beta/(1+\beta)$.
The envelope then stands in for the variance term.
 
\subsection{Participation cap}
 
\begin{figure}[htbp]\centering
\includegraphics[width=0.74\linewidth]{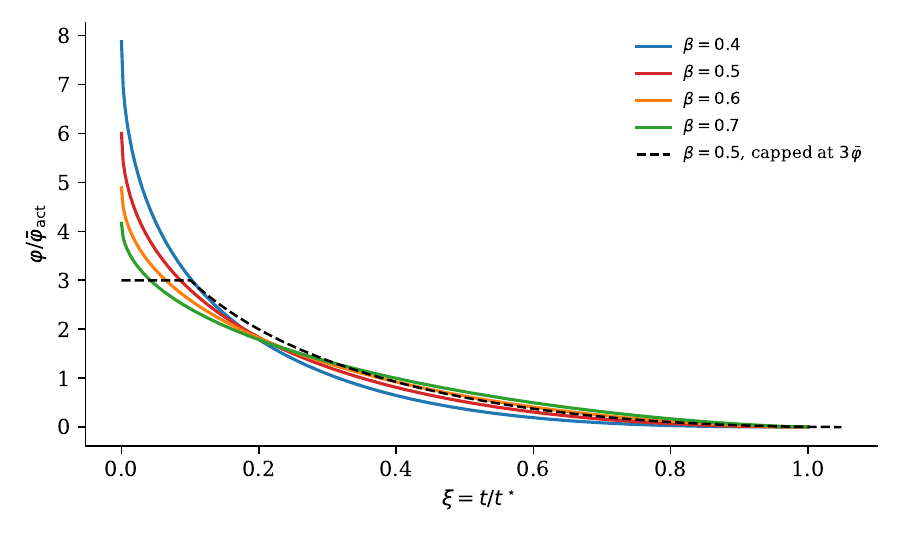}
\caption{Participation $\varphi/\bvphi_{\rm act}$ against $\xi=t/\tstar$ ($H=1/2$) for
several $\beta$, and the capped schedule for $\beta=1/2$ with cap $3\bvphi_{\rm act}$.}
\label{fig:part}
\end{figure}

With a cap $\varphi\le\varphi_{\max}$, KKT applies pointwise:
$\varphi(t)=\min\{\varphi_{\max},[(\lambda-w)_+/\kappa_I(1+\beta)]^{1/\beta}\}$, with
$\lambda$ re-solved from the inventory constraint. The cap binds if and only if
$\varphi_{\max}<\varphi(0)=\bvphi/(K\ustar)$. The schedule then trades at the cap
until $t_1=[(\lambda-\kappa_I(1+\beta)\varphi_{\max}^\beta)/\theta\sigma_1]^{1/H}$ and
follows the interior arc thereafter, finishing later than the uncapped schedule
(Figure~\ref{fig:part}).
 
\subsection{Cost relative to TWAP}
 
Two comparisons are useful, and both are independent of $\theta$, $C$, $X$ and
$\rho$ given the relevant ratio. Against a constant-rate schedule of the \emph{same
duration} $\tstar$, the saving depends only on $(\beta,H)$ (Table~\ref{tab:saving}).
Against a constant-rate schedule over the \emph{whole window}, the saving depends
also on $\ustar$ (Table~\ref{tab:saving2}); it is larger, but mostly because the
benchmark spreads the order over time the envelope penalizes. Therefore, the first comparison is the meaningful one. Both are measured under the envelope; savings under the
price process~\eqref{eq:price} are not implied by this calculation and require
simulation. The saving is in total envelope cost, not in realized impact. At
$\beta=H=\tfrac12$, normalizing TWAP's impact cost to 1, the envelope schedule
pays $1.47$ in impact and only $1.47$ in delay cost\footnote{The equality here (impact = delay cost = 1.47) is not a coincidence: when the deadline does not bind,
compressing the schedule in time at fixed volume scales impact by $s^\beta$
and delay cost by $s^{-H}$, so at the optimum
$\text{impact}:\text{delay}=H:\beta$, which gives equal shares at $\beta=H$.} against TWAP's delay cost of $2.45$. It
therefore shows \emph{higher} arrival-price slippage than TWAP, and whether the
exchange is worthwhile is exactly what $\theta$ encodes.
 
\begin{table}[h]\centering\small
\begin{tabular}{|c|c|c|c|}\toprule
$\beta\backslash H$ & 0.3 & 0.5 & 0.7\\\midrule
0.4 & 14.5 & 16.5 & 16.9\\
0.5 & 12.8 & 14.8 & 15.4\\
0.6 & 11.4 & 13.5 & 14.1\\
0.7 & 10.4 & 12.4 & 13.1\\\bottomrule
\end{tabular}
\caption{Cost saving (\%) relative to TWAP of equal duration.}
\label{tab:saving}
\end{table}
 
\begin{table}[h]\centering\small
\begin{tabular}{|l|c|c|c|c|c|}\toprule
$\ustar$ & 0.10 & 0.25 & 0.50 & 0.75 & 1.00\\\midrule
$\beta=H=0.5$ & 63.5 & 45.6 & 29.5 & 20.4 & 14.8\\
$\beta=0.6,\ H=0.5$ & 63.8 & 45.4 & 28.7 & 19.2 & 13.5\\\bottomrule
\end{tabular}
\caption{Cost saving (\%) relative to TWAP over the whole window.}
\label{tab:saving2}
\end{table}
 
\subsection{Intraday volume}
 
With time-varying $\rho(t)$, define volume time $\dd\tau=\rho(t)\dd t/\bar\rho$.
If, in the diffusive case $H=1/2$, the variance accrues in proportion to volume, and impact is written per
unit of volume, \eqref{eq:J} maps onto itself in volume time. The schedule is then
$v(t)=\rho(t)\,\varphi_{\rm vol}(\tau(t))$. For $H\neq1/2$ the mapping holds only if
the envelope is defined in volume time, which is itself an empirical question.

\section{Properties and Testable Predictions}\label{sec:properties}
 
\begin{corollary}\label{cor:props}
Under~\eqref{eq:envelope}:
\begin{enumerate}
\item[(i)] \emph{Threshold:} $\phi_c=\Theta^{1/\beta}K(\beta,H)$ is
increasing in $H$ for any $\theta$ and $C$, since $\Theta$ does not depend
on $H$ (from 3.8\% to 10.7\% for $H=0.3$ to $0.7$ at $\beta=\tfrac12$,
$\theta=C=1$). For $\theta=C=1$ it is also increasing in $\beta$ (from
5.5\% at $\beta=0.4$ to 11.3\% at $\beta=0.7$, $H=\tfrac12$). The sign of
$\partial u^\star/\partial H$ depends on $\bar\phi/\phi_c$ and is not
determined in general.
\item[(ii)] \emph{Shape:} The normalized schedule~\eqref{eq:shape} depends only on
  $(\beta,H)$. Front-loading, measured by $1/K$, decreases in both: from $8.6$ to
  $3.6$ as $H$ goes from $0.3$ to $0.7$ at $\beta=0.6$, and from $7.9$ to $4.2$ as
  $\beta$ goes from $0.4$ to $0.7$ at $H=1/2$.
\item[(iii)] \emph{Liquidity limits:} As $C\to0$, $\phic\to\infty$ and $\ustar\to0$:
  execute almost immediately. As $C\to\infty$, $\phic\to0$ and the deadline binds.
\item[(iv)] \emph{Volatility:} $\kappa_I$ and $w$ are both proportional to
  $\sigma_1$, so at fixed $T$ and $\theta$ the schedule is volatility-free.
  Volatility enters through $\Tstar$ via $\sigma_1/\mu_f$. This does not hold for
  the alpha reading $w=m$.
\end{enumerate}
\end{corollary}
 
Four predictions can be tested on execution data: (a) Normalized inventory curves
$x(t/t_{\rm end})/X$ should collapse onto~\eqref{eq:trajgen} (eq.~\eqref{eq:traj} for $\beta=1/2$), with a peak-to-average
participation ratio of $1/K(\beta,H)$; given an independent estimate of $H$, this
identifies $\beta$. This prediction does not depend on $\theta$ or on where trading
stops. (b) Across trades, $\iota$ should satisfy \eqref{eq:foc}. This tests the
optimality of the desk's windows rather than the market model. It requires
realized impact and duration from TCA data, together with the signal's own
forecast $\mu_f$ and realization profile ($\tau_\alpha$, $H$), which the desk
already holds. (c) The regressions in Remark~\ref{rem:prior} should locate each stock-day in the
$(a,\nu)$ plane of Proposition~\ref{prop:regimes} (Figure~\ref{fig:regimeplane}),
with $a$ resolved to better than the boundary-layer width $\varepsilon_V$. (d) If $C$ and $\theta$
are common across trades, $\ln\ustar$ against $\ln\bvphi$ should be linear with slope
$\beta/(\beta+H)$.
 
\section{Conclusion}\label{sec:conclusion}
 
When execution is judged by the Sharpe ratio of the trade's realized P\&L, slower
execution lowers risk, whereas urgency comes from alpha realized during the
execution window. The window problem is then organized, to leading order in $T/t_f$, by two distinct signs, the duration
elasticity of impact and the net urgency. Corrections enter through
the exposure-relief elasticity $\varepsilon_V$ and matter only in a thin boundary
layer around $a=0$. Hurst scaling matters mainly because,
under the structural prior $a=\Delta_{IP}=\beta-H$, it can flip the first sign and move a trade
from an interior optimum to the fastest admissible execution. Within the interior
regime the objective is flat, so we report an indifference band.
 
Given the window, a convex programme with concave impact and a per-share cost of
delay yields a closed-form schedule. The schedule minimizes an envelope cost, and its savings are
    savings in that cost; evaluated under the P\&L of Section~3, a
    pessimistic envelope can cost a substantial fraction of the
    Sharpe ratio (Remark~\ref{rem:consistency}).
 Under a Hurst-scaled envelope its normalized
shape depends only on $(\beta,H)$; where it stops within the window is governed by
a single robustness level, $\theta$, for which we give a calibration. 

\paragraph{Open problems:}
First, a local impact model with a fixed impact scale implies a duration
elasticity $a=\beta$, while empirical impact is nearly duration-independent. A
propagator formulation \citep{Gatheral2010} is the natural way to reconcile the
two; convexity survives under no-dynamic-arbitrage, but the KKT condition becomes
an integral equation. Second, the schedules are open-loop; a closed-loop version
that re-solves as fills, prices, and volume arrive would remove the time
inconsistency of any pre-committed plan. Third, the schedule and the window should be chosen jointly, by
    maximizing the Sharpe ratio of (\ref{eq:pnl-density}) over the
    fill density; realized savings under the price process (3),
    rather than under the envelope, would follow from the same
    computation. Fourth, $H$ is treated here as a given input. In practice it is a local, state-dependent exponent, and linking it to observables such as order-book
imbalance, via the propagator exponents of \citet{Bouchaud2004}, would let the
regime classification of Proposition~\ref{prop:regimes} be tracked in real time.

\appendix

\section{Inventory Trajectory for General $(\beta,H)$ } \label{app:traj}

\paragraph{Closed form:}
By~\eqref{eq:shape}, $x(\xi)/X=1-K^{-1}\int_0^\xi(1-s^H)^{1/\beta}\dd s$. The
substitution $z=s^H$, $\dd s=H^{-1}z^{1/H-1}\dd z$ gives
\begin{equation}
  \int_0^\xi(1-s^H)^{1/\beta}\dd s
  =\frac1H\int_0^{\xi^H}z^{\frac1H-1}(1-z)^{\frac1\beta}\dd z
  =\frac1H\,\B\!\left(\xi^H;\tfrac1H,1+\tfrac1\beta\right),
\end{equation}
and dividing by $K=H^{-1}\B(1/H,1+1/\beta)$ gives
\begin{equation}
  \frac{x(\xi)}{X}=1-\Ibeta_{\xi^{H}}\!\left(\tfrac1H,\;1+\tfrac1\beta\right),
  \qquad \xi=t/\tstar\in[0,1],
  \label{eq:trajgen}
\end{equation}
where $\Ibeta_y(p,q)=\B(y;p,q)/\B(p,q)$ is the regularized incomplete beta
function. The normalizing constant $K$ cancels, so the curve has no free
constants: it is the distribution function of a $\mathrm{Beta}(1/H,\,1+1/\beta)$
variable evaluated at $\xi^H$. In SciPy, use
\texttt{1 - scipy.special.betainc(1/H, 1+1/beta, xi**H)}.

\paragraph{Series:}
Expanding $(1-s^H)^{1/\beta}=\sum_{k\ge0}(-1)^k\binom{1/\beta}{k}s^{kH}$ and
integrating term by term gives
\begin{equation}
  \frac{x(\xi)}{X}
  =1-\frac1K\sum_{k\ge0}\frac{(-1)^k\binom{1/\beta}{k}}{kH+1}\,\xi^{kH+1}
  =1-\frac{\xi}{K}\,\Hyp\!\left(-\tfrac1\beta,\tfrac1H;\tfrac1H+1;\xi^H\right),
  \label{eq:series}
\end{equation}
with $\binom{1/\beta}{k}=\frac{(1/\beta)(1/\beta-1)\cdots(1/\beta-k+1)}{k!}$. The
series converges absolutely on $[0,1]$, since
$\binom{1/\beta}{k}=O(k^{-1-1/\beta})$.

\paragraph{Terminating cases:}
If $n=1/\beta$ is a positive integer, the sum stops at $k=n$, with
$K=\sum_{k=0}^n(-1)^k\binom nk/(kH+1)=H^{-1}\B(1/H,n+1)$.
\begin{itemize}
\item $\beta=1$ ($n=1$): $K=\dfrac{H}{H+1}$ and
  \begin{equation}
    \frac{x(\xi)}{X}=1-\frac{H+1}{H}\Bigl[\xi-\frac{\xi^{H+1}}{H+1}\Bigr].
  \end{equation}
\item $\beta=1/2$ ($n=2$): $K=\dfrac{2H^2}{(H+1)(2H+1)}$ and
  \begin{equation}
    \frac{x(\xi)}{X}=1-\frac{(H+1)(2H+1)}{2H^2}
    \Bigl[\xi-\frac{2\,\xi^{H+1}}{H+1}+\frac{\xi^{2H+1}}{2H+1}\Bigr],
  \end{equation}
  the factored form of~\eqref{eq:traj}.
\item $\beta=1/3$ ($n=3$): four terms, with coefficients $1,-3,3,-1$ on
  $\xi,\ \xi^{H+1}/(H+1),\ \xi^{2H+1}/(2H+1),\ \xi^{3H+1}/(3H+1)$, and
  $K=\dfrac{6H^3}{(H+1)(2H+1)(3H+1)}$.
\end{itemize}

\paragraph{Non-integer $1/\beta$:}
For $\beta=0.6$ ($1/\beta=5/3$) the series~\eqref{eq:series} is infinite, and
the incomplete-beta form~\eqref{eq:trajgen} is the finite closed form.

\end{document}